# HEAT FLOWS AND ENERGETIC BEHAVIOR OF A TELECOMMUNICATION RADIO BASE STATION

Antonio Petraglia[1], Antonio Spagnuolo[2], Carmela Vetromile[2], Antonio D'Onofrio[1], Carmine Lubritto[2]

[1]Department of Mathematics and Physics (DMF), Second University of Naples, Viale A. Lincoln, 54, I - 81100 Caserta – ITALY

[2]Department of Environmental Science and Technology (DiSTABiF), Second University of Naples, Via Vivaldi, 43, I - 81100 Caserta – ITALY

Keywords
**Power consumption; Base transceiver stations; Sustainable development; Thermal balance of a shelter; Energy savings**

Highlights
- A heat flow model for a telecommunication shelter was created.
- Successful matching of the model and real-life cases was obtained.
- The model addresses how each parameter affects energy consumption.
- Reasonably savings of up to 30% of consumed energy can be expected.

Abstract

This paper shows a study on energetic consumption of Base Transceiver Stations (BTSs) for mobile communication, related to conditioning functions. An energetic "thermal model" of a telecommunication station is proposed and studied. The results have been validated with a BTS in central Italy, showing good agreement. Findings show a substantial high internal-external temperature difference in the containing shelter, particularly during daytime and warm months, due to sources of heat (equipment, external temperature and sun radiation) and to the difficulty in spread the warmth out. The necessity to keep the operating temperatures within a given range for the correct functioning of the electronic equipment requires the use of conditioning setups, and this significantly increases the energetic demand of the whole system. The analysis of thermal flows across the shelter can help to gather further data on its temperature behavior and to devise practical measures to lower the power demand, while keeping the operating parameters in the suggested ranges. The investigation of some operating parameters of the equipment and of the shelter, such as threshold set-points, air vent area, external wall transmittance and reflectivity, suggests annual energy savings between 10% and 30%.

## 1. Introduction

The technological development of the telecommunications sector goes hand in hand with the gradual increase of its energy consumption. In recent years, the focus has been to maximize the savings and use energy efficiently [1][2][3][4][5].

The BTS have a high-energy consumption due to the transmission equipment and the air conditioning of the shelter (the "building" that contains the transmission apparatuses). The energy consumption of the air conditioning system accounts for about 40-45% of that of the entire system [6][7].

Within the shelter, the thermo-hygrometric parameters, such as temperature and humidity, can be rather different from those recorded on the outside, due to the presence of many devices that dissipate energy as heat. However, they have strict operating parameters, and thus the use of machinery for air conditioning is indispensable to protect against dew and high temperatures.

The conditioning equipment must perform its task whilst taking into account energy savings [8].

Free cooling is a good opportunity for energy saving in air conditioning systems, due to its high efficiency. In fact, it is used when outside air is cool enough to be employed as a cooling medium [9]. It is usually of two types: air-side and water-side free cooling. The first introduces air into the shelter, whatever the conditions [10][11]; water-side, through cooling towers or dry coolers, is used to aid the outdoor air compressors in cooling the water supplied to the chiller system [12]. The use of these devices assists the air conditioning [13] and allows for good energy savings, especially during the transition periods (April, May, October) [9].

For the purpose of energy saving, it is also useful to set up correct management of the cooling systems, both through the diagnosis of abnormal energy consumption, due to errors in sensors or in cooling devices [14]; and through the correct setting of thresholds for the use of air conditioning systems [15], whereby, as shown in [16], it is possible to save up to 25% of cooling energy.

Other key elements for the study of thermo-hygrometric parameters within the shelter are the building materials.

They must be chosen in relation to the climatic conditions of the area where the stations are located [17][18].

In this paper, we present a study of the temperature behavior of the shelters, concentrating in particular on the heat fluxes connected to them. This investigation is directed towards two goals: i) study of the energetic fluxes in the shelters; ii) optimization and dynamic control of the air conditioning systems. This paper models the power consumption associated with the conditioning of a station, studied through an algorithm that allows to outline the shelter as a thermodynamic system with heat fluxes (due to conduction, convection, and radiation) both towards and away from the building. The parameters that influence the thermal load of the shelter are complex and interconnected. The study and comprehension of the relationships between them is useful to develop systems that are able to simulate and predict the heat fluxes inside the shelter, in order to optimize energy consumption management [10][18][19][20]. To this end, hybrid energy systems with smart use of the cooling equipment have also been suggested [21].

With the algorithm, it is possible to change a series of parameters that characterize the shelter (the size of the free cooling window, the temperature thresholds of the conditioning system thermostats, the reflectivity of the building, the thickness and building materials for the walls and so forth.).

Therefore it is possible to study how each parameter affects energy consumption of the station and to evaluate the improvements that can decrease it.

The results were validated by monitoring a base station owned by the telephone service provider Wind, located in central Italy. They are consistent with results obtained elsewhere, both in Italy [6] and in other climates [10][18], and also in situations where a similar method was used to study energy saving when a thermosyphon heat exchanger is used [22].

## 2. Thermal balance of a shelter

A "thermal model" of shelters for telecommunication, including internal apparatuses and external contributions was created, to study its thermal balance. Therefore, it was possible to perform simulations, varying the factors that characterize the energy behavior of a BTS, to address the most effective parameters to lower energy consumption, recalling that the thermodynamic quantities (temperature, humidity, etc.) of a shelter for telecommunication can show a much wider range of values compared to places where there is human presence [7].

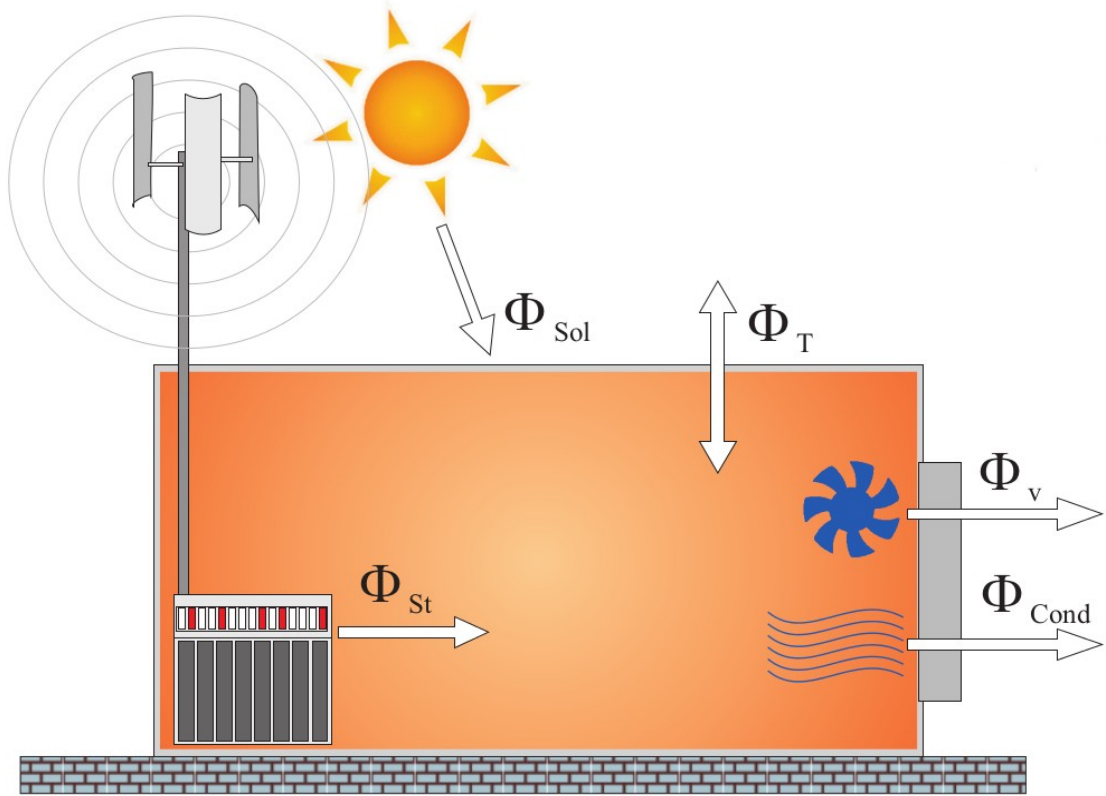

Fig. 1. Schematic representation of the energetic fluxes in the shelter (see text).

Assuming that the system is in thermal equilibrium with the external environment, the total heat flow in the system equals zero:

$$\Phi_T + \Phi_V + \Phi_{sol} + \Phi_{Cond} + \Phi_{St} = 0 \tag{1}$$

where $\Phi_T$ is the heat flux through the walls, $\Phi_V$ is the heat flux due to ventilation (also called free cooling), $\Phi_{Sol}$ is the heat flux due to solar radiation, $\Phi_{Cond}$ is the flux due to air conditioning and $\Phi_{St}$ is the heat flux released by the instrumentation (internal flow). The sign convention used here considers the entering heat flow as positive and expelled heat as negative. The energetic fluxes of a telecommunication shelter are shown schematically in Fig.1.

As simplifying assumptions, the temperature is supposed to be uniform through the entire shelter (lumped temperature model) and all the parameters are supposed to be homogeneous and constant. Identically, the external temperature is supposed to be homogeneous. Higher order effects, such as wind on the external wall, are not considered; moreover, thermal properties of the walls are supposed to be homogeneous.

The flux contributions were modeled as follows [23][24]:

*A. heat transfer through the walls.*

This contribution is due to conduction and depends on the temperature difference between the inside and the outside of the shelter. Heath enters the system during the hottest days and sinks at night and on cool days. For this reason, this flow may be either positive or negative depending on the sign of the temperature difference:

$$\Phi_T (W) = A_{tot} * U * (T_{out} - T_{in}) \tag{2}$$

where $U$ is the transmittance of the wall, $T_{in}$ and $T_{out}$ are the mean internal and external temperatures, respectively, and $A_{tot}$ is the sum of the areas of the four side walls and the ceiling. The floor was neglected since there is no noticeable heat flow through it.

*B. Heat flow due to ventilation (free cooling).*
This input is due to apparatuses that expel hot air outside and pump the external air in. The advantage of these systems is primarily the use of low external temperatures to save energy during cold periods of the day/year. The only consumed electrical energy is the one used to power the fans. On cold days, this is the most efficient cooling method.
The heat flux due to free cooling is:

$$\Phi_V (W) = - H_v * V_a * \Delta T * A_{FIN} \tag{3}$$

where $H_v$ is the volumetric heat capacity of the air. It depends on air density, $\rho_a = 1.2$ kg/m$^3$, and on the specific heat of air at constant pressure $c_a = 1000$ J/(kg K), so that $H_v = \rho_a * c_a = 1200$ J/(m$^3$ K).
$V_a$ is the velocity of the incoming air; $\Delta T$ is the temperature difference between the inside and the outside. $A_{FIN}$ is the area of the window of the free cooling system.
The negative sign indicates that this type of flow disperses energy towards the external environment. Practically, it is activated with pre-programmed thermostats.

*C. Heat flow from solar radiation.*
The sun radiation is accounted as:

$$\Phi_{sol} (W) = G * A * (1 - \rho) \tag{4}$$

where $G$ is the solar power that arrives on one m$^2$ of soil and $(1 - \rho)$ is the absorptivity of the walls of the shelter. $A$ is the actual area affected by the solar radiation. It is calculated, for each exposed surface, as:

$$A = A_{SUP} \cos(\theta); \tag{5}$$

here, consider the direct solar radiation striking the surface on sunny days with an angle θ between the sun position and the normal of the surface. In the following, the position of the sun (altitude and azimuth) was calculated through an algorithm taking as input the chosen day and geographic position of the site. The sun position was calculated from an algorithm derived from [25], and shown at the site http://www.srrb.noaa.gov/index.html (in document solareqns.pdf). The error noted in http://www.mail-archive.com/sundial@uni-koeln.de/msg01050.html was fixed. The maximum error declared by Spencer was 3 minutes. The algorithm was tested with a number of online sun position calculators (at http://www.esrl.noaa.gov/gmd/grad/solcalc/ and http://stackoverflow.com/questions/8708048/position-of-the-sun-given-time-of-day-latitude-and-longitude) with differences lower than a half degree, which is acceptable for this analysis. This is confirmed by the study of the algorithm made in [26] which entails a 0.28 degree of maximum error.

The effect of the clouds was not considered due to the variability of this contribution; its account should be considered for finer yearly consumption calculations. Instead, this choice gives the worst-case scenario (i.e. higher conditioning consumes), especially on hot summer days, because it maximizes the solar heat flux.

*D. Heat input from the equipment.*
This impact is due to unwanted heath dispersion of all the electrical apparatuses; it depends on the transmission technology used. In this case (analyzing the consumption for the BTS FR001, used for testing the good quality of the algorithm), this value is 2/3 of 108 kWh/d [6]; consequently, the average heat flow *$\Phi st$ = 3000 W* for the station studied was considered. It is possible to consider a time variation of this value during the day/week, to obtain a closer match between simulations and data. However, it was chosen to leave it constant in time, which is an acceptable first approximation.

*E. Air conditioning.*
$\Phi_{Cond}$ is the flux due to air conditioning and it is used to balance the other fluxes, to keep the system in the temperature range allowed for the correct functioning of the instrumentation. Multi thermostatic temperatures for multilevel conditioning power (as in real-life cases) were considered.

## 3. Characteristics of the base station monitored

The monitored station is located in the municipality of Frosinone in central Italy. In Table 1 are reported the characteristics of the station, the sampling period and the type of equipment for the transmission.

| SITE CODE | FR001 | | |
|---|---|---|---|
| **Station typology** | SHELTER | | |
| **Sampling periods** | 24/09/2013 – 02/10/2013<br>04/12/2013 – 11/12/2013 | | |
| **GSM Transmitters** | 6 | | |
| **DCS Transmitters** | 14 | | |
| **UMTS Transmitters** | 3 | | |
| *Station apparatuses* | | | |
| **Supply** | 380 V ENEL | | |
| **Power station** | Emerson Ulisse 4 | | |
| **Rectifier** | BML440051/1 | | |
| **Conditioning + Free cooling (x2)** | AEA EL70CWVD | | |
| **Base Transceiver System** | GU3900 HUAWEI | | |
| *Dimensions* | | | |
| | Height | Width | Depth |

| | | | | |
|---|---|---|---|---|
| **Shelter (m)** | | 2.45 | 3.30 | 2.20 |
| **Equipment (m)** | **Condition./Free Cool.** | 2.00 | 0.63 | 0.20 |
| | **Conditioning grate** | 0.29 | 0.55 | |
| | **Ventilation grate** | 0.20 | 0.20 | |
| | **Emerson network** | 1.82 | 0.60 | 0.43 |
| | **BTS cabinet** | 1.84 | 0.59 | 0.45 |
| | **Main Distribution Frame** | 1.60 | 0.26 | 0.30 |

Table 1. Characteristics of the shelter and equipment of the station under study.

The internal cooling of the station consists of two free cooling (FC) systems and two air conditioning (CDZ) systems with different threshold temperature values.
The on / off of the air conditioner and free cooling is controlled by a PLC (Programmable Logic Controller), with preset "ON" thresholds and associated hysteresis values. The latter represents the temperature drop needed (after reaching the "ON" threshold) so that the cooling machine turns itself off. Table 2 shows the thresholds of switch-on and the related hysteresis values.

| | | |
|---|---|---|
| **FC1** | **THRESHOLD (°C)** | 24 |
| | **HYSTERESIS** | 1 |
| **FC2** | **THRESHOLD (°C)** | 25 |
| | **HYSTERESIS** | 2 |
| **CDZ 1** | **THRESHOLD (°C)** | 27 |
| | **HYSTERESIS** | 1 |
| **CDZ 2** | **THRESHOLD (°C)** | 28 |
| | **HYSTERESIS** | 1 |

Table 2 Thresholds of switch-on and the related hysteresis values for the considered BTS. FC1 and FC2 indicate the first and second free cooling system; CDZ1 and CDZ2 indicate the first and second conditioning system.

Besides the ON / OFF thresholds presented in table 2, the PLC activates the free cooling if the outdoor temperature is lower than 24 °C, and if the difference between indoor and outdoor temperature is higher than 2 °C.

## 4. Materials and methods

In the monitoring campaign, the following operating parameters were measured:
• Energy consumption (kWh)
• Indoor Station Temperature (°C)
• Outdoor Station Temperature (°C)
• Global Solar Radiation (W/m$^2$)

The measurements were acquired via:

- **LSI LASTEM - Elog:** a data-logger for environmental applications, to which the signals from the various sensors used for the measurements were fed. Specifically:

    **Pt100 probe:** it provides a temperature reading. Apart from the Pt100, the LSI LASTEM is equipped with its own inner thermocouples. This made possible to read both the internal and external temperatures of the station.

    **Phonometer Smart Sensor Model AR814**: noise was measured to gain deeper understanding of the dynamics of the ON / OFF status of the air conditioning equipment.

    **Radiometer:** this permits the measurement of global radiation.

- **Wireless Monitor CM160:** data-logger that allows to record electricity consumption in real time. The monitor acquires data sent from a transmitter to which three amperometric clamps (one for each phase of the electrical panel of the station) are connected.

Simulations were performed with scripts written in the R language [27]. The used algorithm is shown in Fig. 2. The variables are:

- $T_i$ = internal temperature of the shelter;
- $T_e$ = temperature of the environment outside the shelter;
- $T_{COND}$ = threshold temperature of the conditioner (different for every conditioner);
- $T_{hyst}$ = hysteresis (in general different for every conditioner and free cooler);
- $T_{FC}$ = threshold temperature of the free cooler (different for every free cooler);
- $T_{e\_max}$ = maximum external temperature to have free cooling on;
- $\Delta T_{min}$ = minimal external/internal temperature difference to have free cooling on.

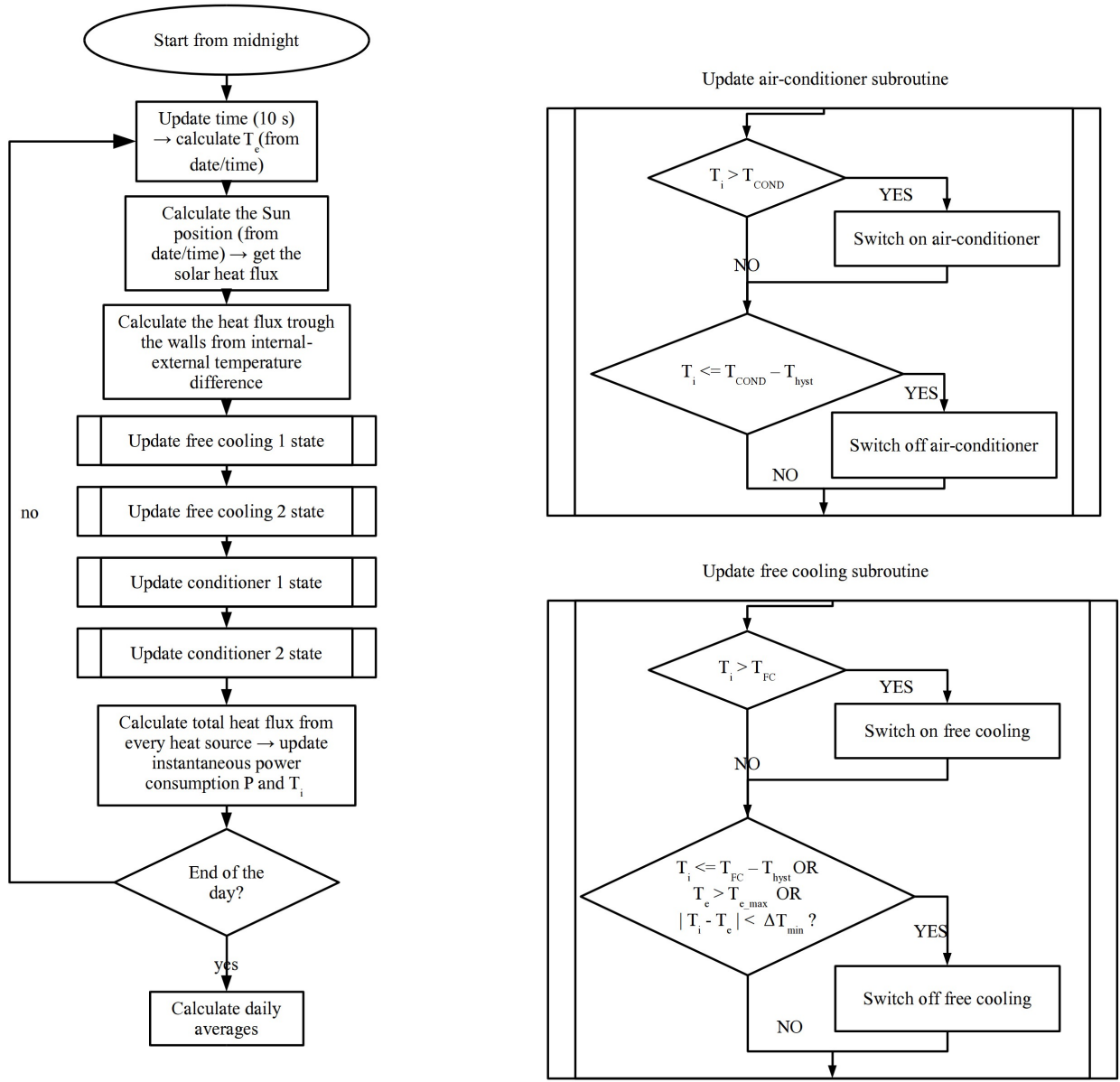

Fig. 2. Schema of the algorithm used for simulations. The variables are explained in the text. The algorithm calculates the internal temperature and the power consumption every 10 seconds during the whole day, by considering the total heat flux balance, then calculates the daily averages.

## 5. Simulations

Simulations were carried out with the following input data, based on real data taken from the station FR001:
- $V_s$ (volume of the Shelter) = 18 m³
- effective area of the shelter, disregarding the floor and ceiling: $A_{tot} = 30\ m^2$
- Maximum cooling power of conditioners = (2 x 6000) W
- Free cooling air flow = 400 m³/h

- $\rho$ (Reflectivity of the external walls) = 0.7975
- Reflectivity of the external nearby ground = 0.2
- Area of the free cooling window = 0.1 m$^2$
- free cooling air velocity = 0.9 m/s
- The outdoor air temperature was chosen from the measurement campaign. The external temperature was modeled with a sinusoidal curve with a one-day period, oscillating between the minimum and maximum of the daily averaged temperatures of the considered month. The minimum was chosen at 03:00 AM.
- The solar contribution to the heat flow was calculated with an algorithm which accounts for the position of the sun for each day and geographic position (see above). The sun position was calculated for every time step of the simulation.
- The wall considered for the calculations is formed, starting from the outside, by a sheet of aluminum or galvanized steel a few mm thick, an intermediate layer of polyurethane foam with a conductivity $K$ between 0.02 and 0.06 W/(m$^2$ °K) and an inner layer of sheet aluminum. For conductivity, only the polyurethane foam was considered, because of its predominance on aluminum in the transmission of heat ($K = 0.021$ W/(m$^2$ °K)) and a wall thickness L = 0.05 m. The transmittance is thus U = 0.7 W/(m$^2$ K). Using these parameters, the heath flow through the walls is very low. Indeed, for a difference between external and internal temperature of 10 °C, $\Phi T = A_{tot} * U * \Delta T = 30 * 0.7 * 10 = 210$ W. This is because the shelters are built with highly insulated walls. This is a good choice only in warm climates, where it is better to insulate the apparatuses from the external temperature.
- The electric equipment utilized for transmission and functioning of the entire apparatus consumes a considerable amount of power with a total heat flow $\Phi_{st} = 3000$ W.
- The conditioning apparatuses have a power consumption of 500 W for each free cooling and 2500 W for each air conditioner.
- The thermal capacity of the apparatuses and internal structures was modeled with an additional air mass of 600 kg. This is half of the effective metal mass, because the specific heat of metals is about half the specific heat of the air.

## 6. Results and discussion

Simulations were performed for 24-hour periods with 10-second time interval steps (as shown in Fig. 2).
Fig. 3 shows the sampling days on the abscissa, the energy daily consumption (bars), both simulated and measured, on the main ordinate, and average MIN and MAX external temperatures (lines) on the secondary ordinate.

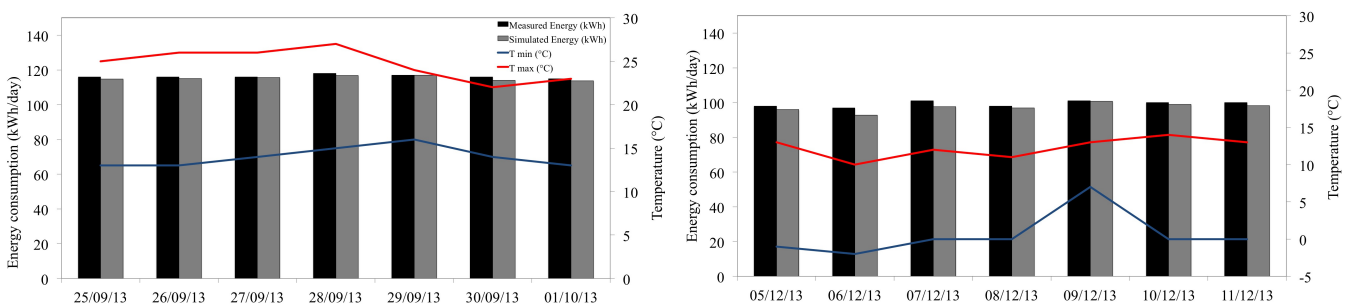

Fig. 3. Measured and simulated BTS daily energy consumption and min/max external temperature for the two weeks under study: 25/9/2013-1/10/2013 and 5/12/2013-11/12/2013.

The weeks of sampling represent two distinctly different periods, concerning environmental conditions; in fact, the week of September is typical of the environmental conditions of early autumn (Italian climate) with rather mild temperatures; the one-week sampling in December shows lower temperatures. There is agreement both in the week of September and December: in each case, the deviation of the simulated value does not exceed 5% of the measured value.

The instantaneous power consumption of the station FR001 during two days is shown in fig. 4. The abrupt changes in power consumption correspond to the on/off of the free cooling and the conditioner, depending on the internal/external temperature status.

It is possible to see that similar on/off behavior exists in experiments and simulations, both during hot and cold days.

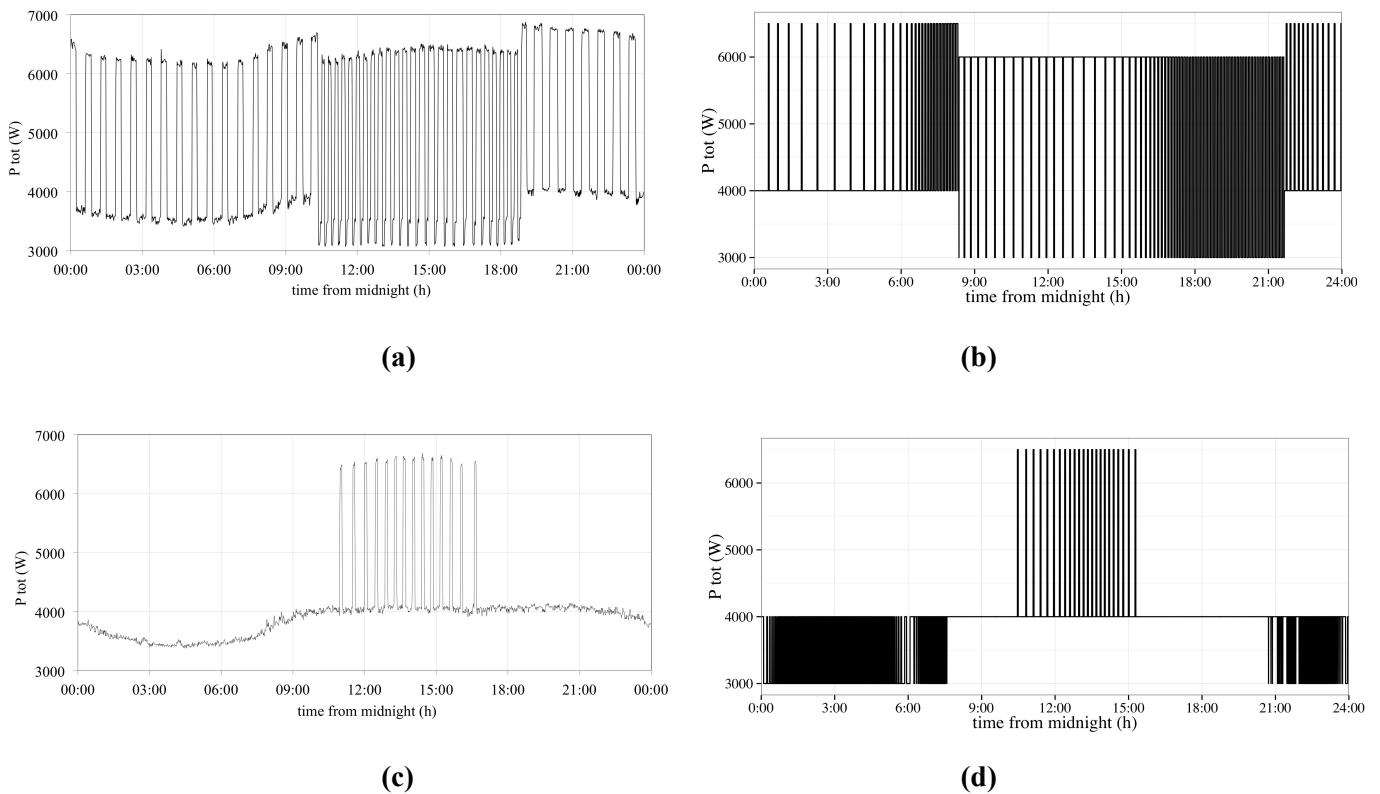

Fig. 4. Instantaneous power consumption in two days: 28 September 2013: (a) measured, (b) simulated; and 6 December 2013: (c) measured, (d) simulated.

The trend of September is typical of the warm periods, with a night and evening consumption characterized by free cooling always ON, the ON/OFF alternation of the air conditioner and a consumption, during the day, characterized by the ON/OFF of the air conditioners and free cooling always OFF, because the external threshold temperature for the deactivation is reached. The December trend is typical of colder months, with free cooling ON throughout the day (not perfectly reproduced by the simulation algorithm) and the ON/OFF of air conditioners only during the hottest hours of the day.

## 6.1 Monthly energy consumption

The average total daily energy consumption was simulated for every month of the year. For the simulations, the central day of each month was chosen with the average MIN and MAX measured temperatures of the month.

The simulation results are shown in Fig. 5.

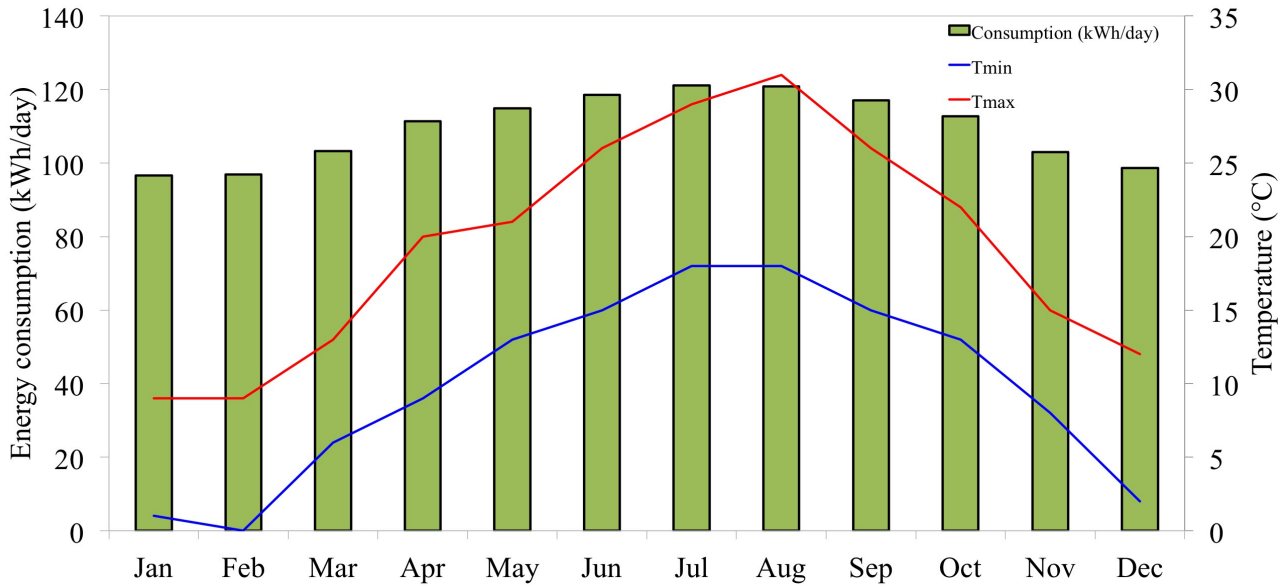

Fig. 5. Average daily total energy consumption, calculated for each month.

The abscissa shows the months, while the main ordinate the daily energy consumption (bars) and the secondary ordinate the min and max monthly temperatures (lines).

From the graph, it can be seen that the energy consumption follows temperature changes over the year. In fact, there is a higher consumption during the hottest months of the year, due to the use of air-conditioning systems, to the higher temperatures and to the increased solar radiation during spring and summer periods. The difference between the warm months and cold months can reach 20%.

From these simulations, it is also possible to project the annual consumption of the station by multiplying the results obtained from the simulations per day by the respective number of days per month. Thus, an annual energy consumption of 40,000 kWh was estimated. This result is consistent with the results obtained in [6] where the energy consumption of BSs was calculated from statistical data of annual consumption for 95 BSs on the Italian territory.

**6.2 Temperature set-point regulation**

The research also focused on the possibility of energy savings potentially accessible through the setting of the activation thresholds of free cooling and air conditioners.
The simulations were carried out by increasing the set-points of the apparatuses of conditioning.
Fig. 6 shows the evolution of annual energy consumption in relation to the increase of the FC-CDZ thresholds. As can be seen, it is possible to save about 10 % for an increase of 10 °C in the FC-CDZ activation thresholds, with a quasi-linear correlation between the two variables.

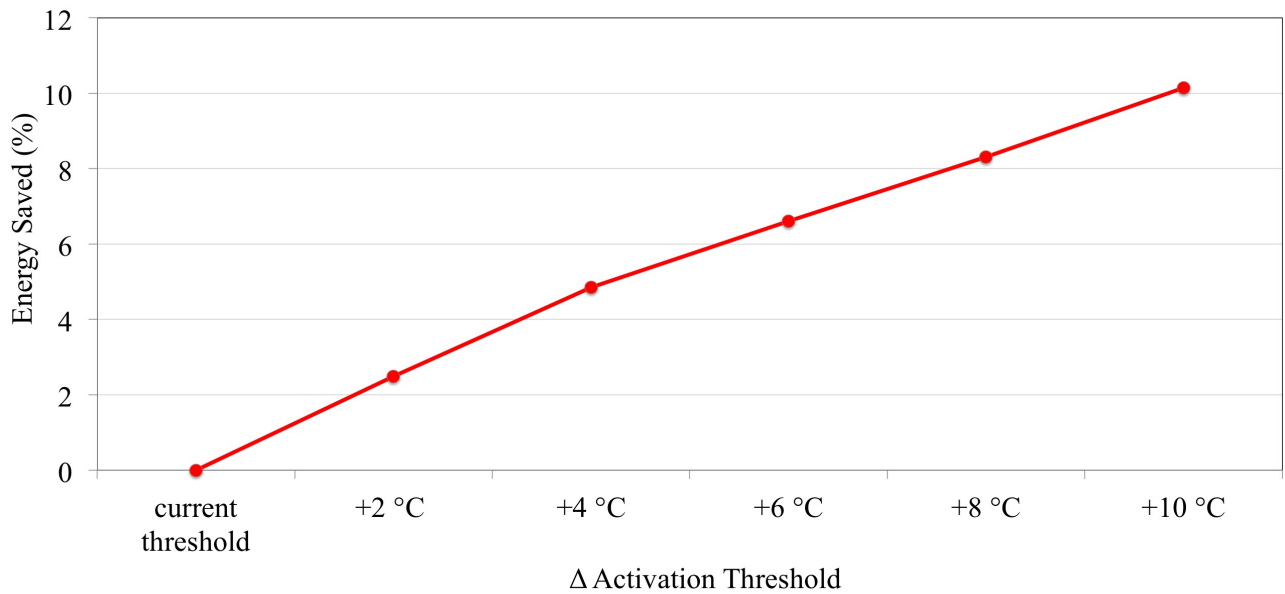

Fig. 6. Yearly energy consumption versus air conditioning set-point.

The months during which there are the greater energy savings are the colder ones; this is due to the possibility of having the conditioning apparatuses OFF for longer times, by increasing the thresholds of activation and thus taking advantage of the low external temperatures. This is shown in Fig. 7, which presents the trend of the average daily energy consumption versus the months, and the air conditioning temperature set points.

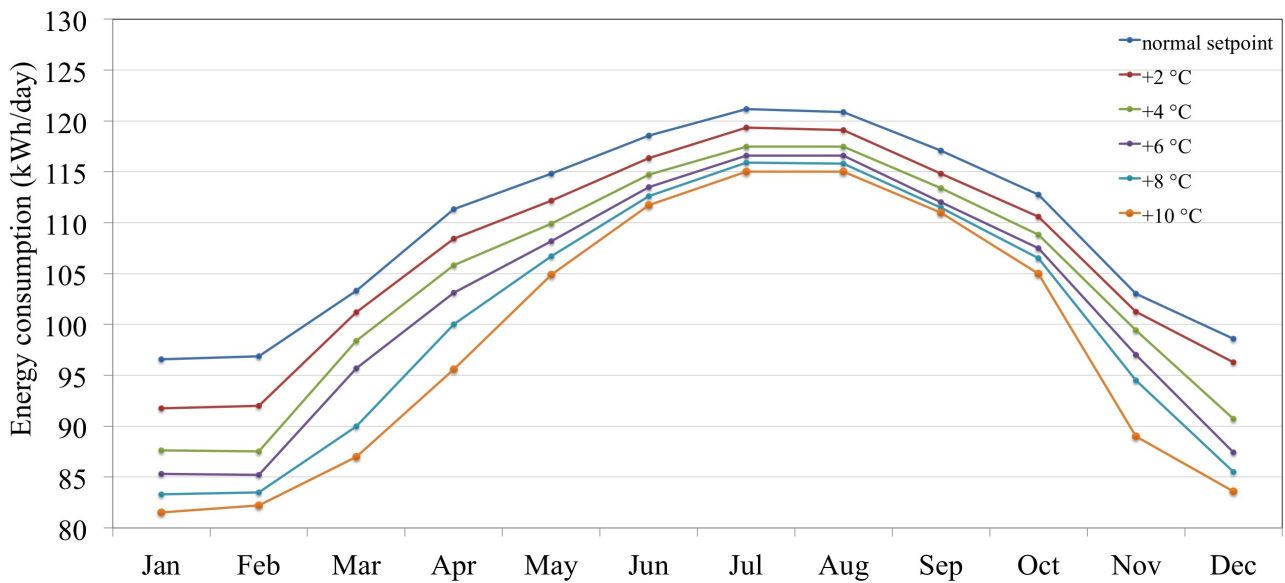

Fig. 7. Average daily energy consumption versus the months and the temperature set-points.

Naturally, the activation thresholds should be below a definite value, to satisfy the equipment operating parameters within the shelter and to avoid incurring in damage caused by high temperatures.

## 6.3 Air vent surface

Air vent surface affects the flow rate and the cooling capacity of air conditioning systems: larger areas correspond to a greater flow rate. Fig. 8 shows the annual amount of saved energy through an increase of the air vent surface of the equipment. Remember that the surface of the shelter under study is 0.10 m², and by increasing it to 0.15 m², it is possible to save about 2300 kWh per year (6 % of the total).

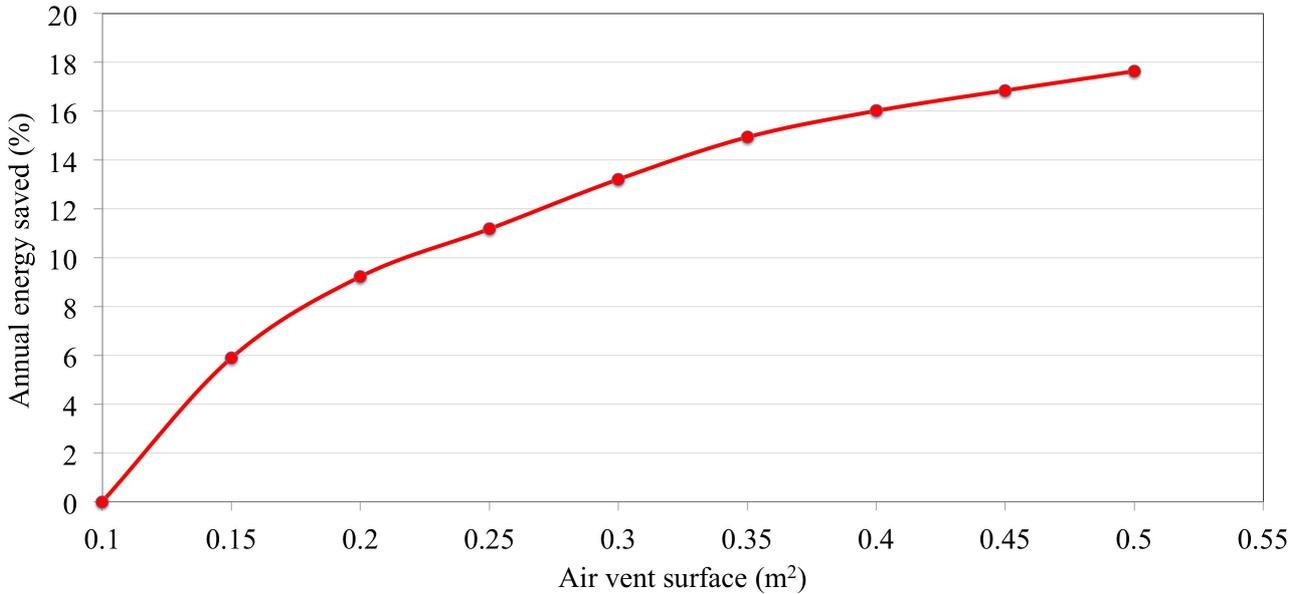

Fig. 8. Annual amount of saved energy through an increase of the air vent surface of the free cooling.

In this case, the months with higher savings are also those with lower temperatures, because it is possible to use a larger amount of cold air to cool the interior of the shelter.

## 6.4 Thermal transmittance

Thermal transmittance is an important parameter, especially for some climatic conditions: the right relationship between insulation and thermal conductivity is essential to avoid energy waste from conditioning [18].
Fig. 9 shows the annual energy savings versus the thermal transmittance of the shelter walls. Starting from a transmittance of 0.7 W/m²*K (which is the value corresponding to the analyzed shelter, see above), the energy savings increase with the transmittance, although slightly, to about 10 % of the total, with a transmittance of 5 W/m²*K.

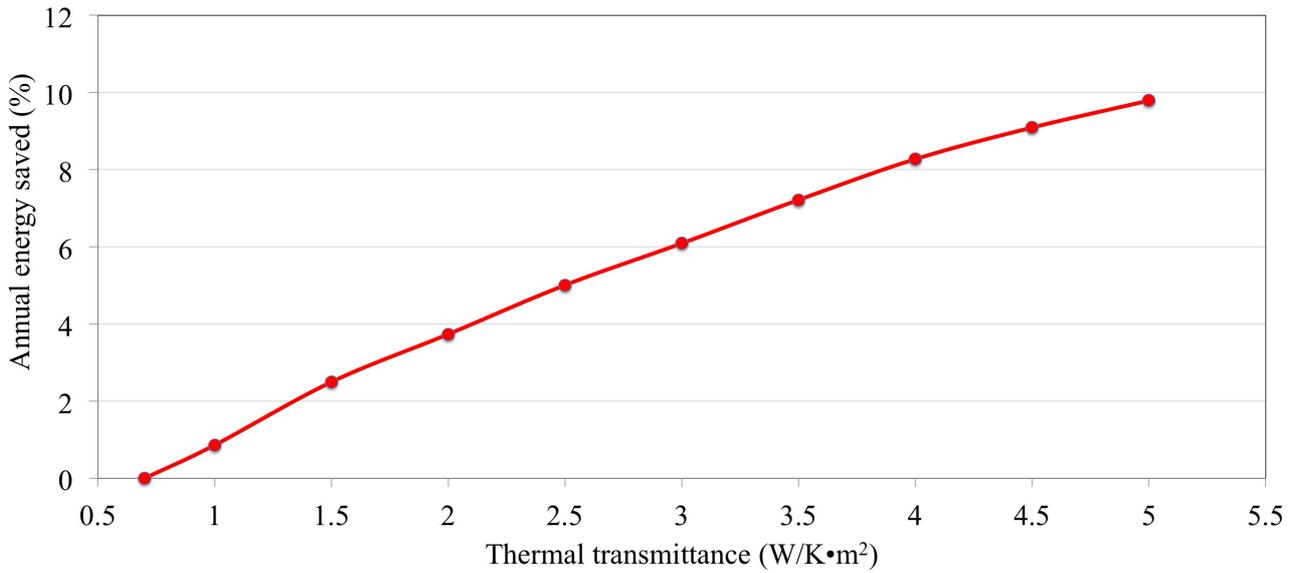

Fig. 9. annual energy savings versus the thermal transmittance of the shelter walls.

This mainly arise from the particular climatic conditions of the station site; in fact, it is characterized by mild winters and hot summers: the increase in transmittance in winter helps to cool the interior of the shelter, while it is almost ineffective in summer, when the heath flux through the wall is negligible compared to the total heath.
This situation is depicted in Fig. 10, where the average daily consumptions versus the different thermal transmittances, for each month of the year, are shown.

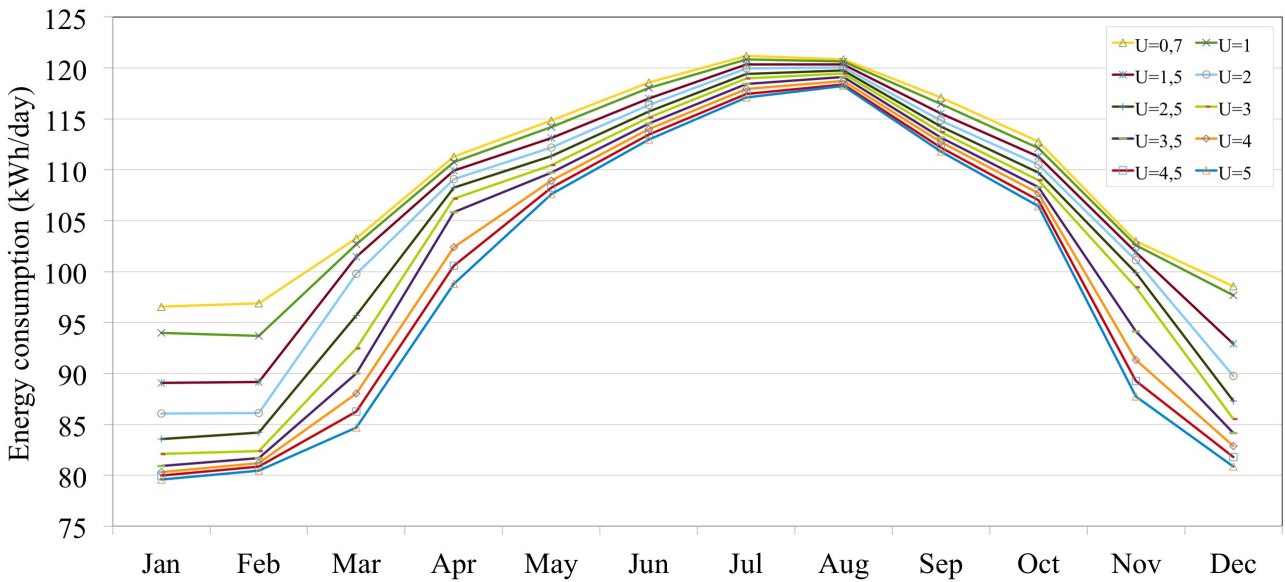

Fig. 10. Average daily consumptions versus the different thermal transmittances, for each months of the year.

The saving is evident in the low-temperature months, but it becomes almost negligible as the external temperature increases.

**6.5 External walls reflectivity**

Finally, this research also focused on the influence of the external walls reflectivity on the energy consumption of the station. To perform the simulations, RAL (Reichsausschuss für Lieferbedingungen) values were used. They are utilized for information defining standard colors for paint and coatings, and are the most popular Central European Color Standard used today. A different reflectivity is associated to any RAL color. This fieldwork employed the colors in Tab. **3.**

| CODE | NAME | Reflectivity value (ρ) |
|---|---|---|
| RAL 9004 | signal black | 0.01 |
| RAL 5010 | gentian blue | 0.05 |
| RAL 5023 | distant blue | 0.12 |
| RAL 6018 | yellow green | 0.25 |
| RAL 7001 | silver grey | 0.31 |
| RAL 7032 | pebble grey | 0.44 |
| RAL 1021 | cadmium yellow | 0.52 |
| RAL 9018 | papyrus white | 0.61 |
| RAL 9001 | cream | 0.78 |
| RAL 9016 | traffic white | 0.85 |
| Reflective paint | ** | 0.95 |

Table. 3. RAL color codes and names, and relative reflectivity values.

Shelters have a RAL code of 9001, with a reflectivity of 0.78. Fig. 11 shows the average annual consumption of the shelter versus reflectivity. As it is possible to see, the power consumption varies from approximately 53300 kWh per year, for a shelter with RAL 5023 (black), to 40000kWh of a normal shelter color (RAL 9001), until reaching an average annual consumption of 39000 kWh for a shelter painted with ρ = 0.95 paint.

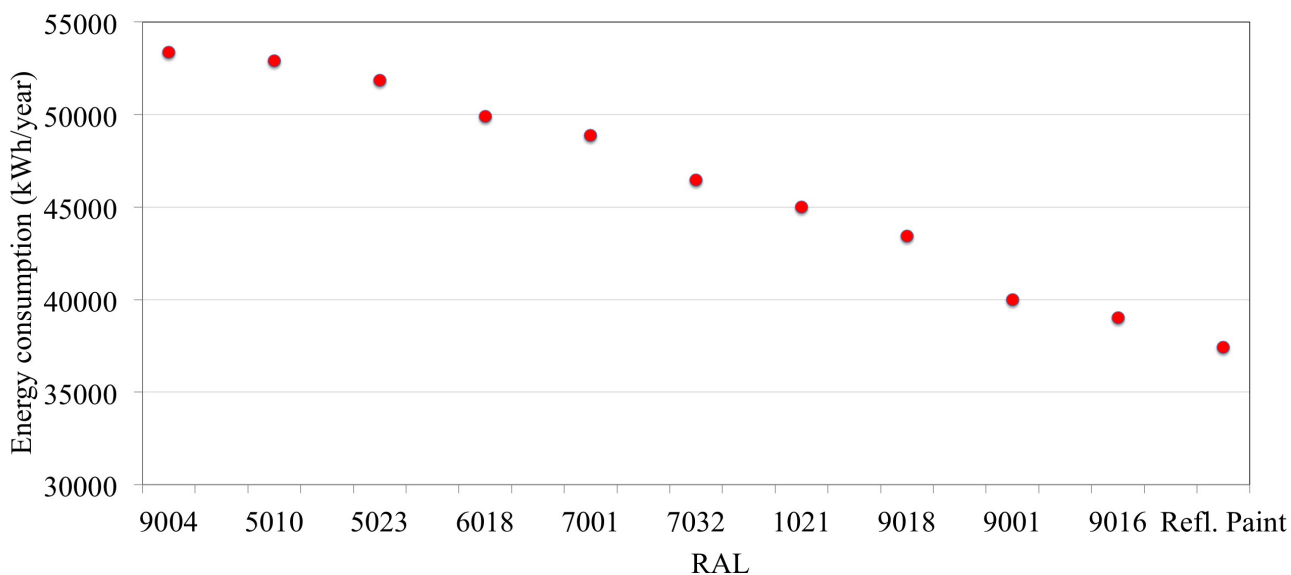

Fig. 11. Average annual consumption of the shelter versus reflectivity of the external walls.

It is therefore deducible that the reflectivity has a higher weight in months with the greatest quantity of solar radiation. In Fig. 12, it is possible to note that to very low reflectivity values (i.e. closer to one) correspond a very high average daily consumption, especially in warm months, when the solar radiation contribution becomes significant.

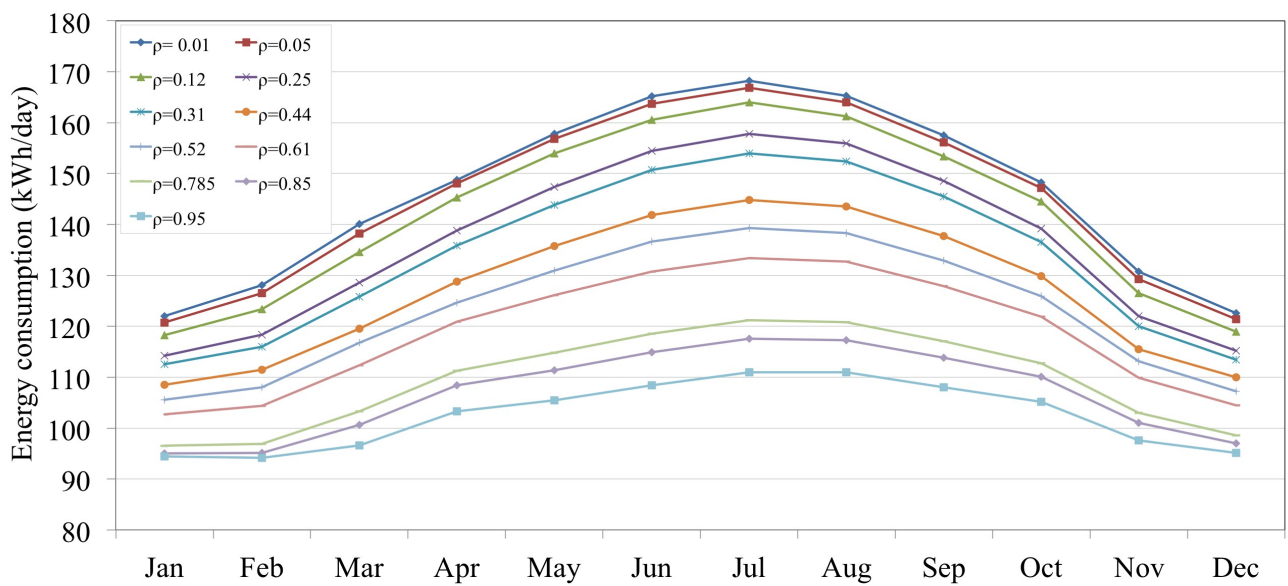

Fig. 12. Average daily consumptions versus reflectivity of the shelter external walls.

## 7. Conclusions

In this paper, an evaluation method for the energetic savings of a BTS has been shown. It is based on a "thermal model" that correlates the power consumption of the BTS with its environmental, geometric and operating parameters.

The algorithm takes into account a reference BTS located in the municipality of Frosinone (central Italy), of which the operating parameters, the energy consumption and the environmental conditions of the sampling period were monitored.

The results led to a good match between the simulated and the actual daily consumption, and between the trends in energy consumption during the day and the annual energy consumption, which for the station under consideration is 40,000 kWh.

The full year simulation showed total savings between 10% and 30% of the annual consumed energy. Moreover, there was a 20% difference in energy consumption between the hot and the cold months.

The increase of the temperature set point has revealed 2% of energy savings for every increase of 2°C (with the coldest months being those with greater savings.). This is interesting because an increase of a few degrees from the actual state can reasonably be foreseen, also because the set point can be easily changed with simply a one-time intervention.

The surface of air vent of the free cooling equipment bears heavily on energy savings, with about 2,300 kWh / year (6% of the total) for an increase of 0.05 m$^2$; on the other hand, by increasing the surface area of 0.4 m$^2$, the resulting annual energy savings is about 7,000 kWh (18% of the total).

Increasing the thermal transmittance of the shelter walls does not give large variations in annual energy consumption, if not at high values, while it has some evident effects in the colder months.

The reflectivity of the external walls of the shelter were also studied, starting from very low reflectivity (dark colors), up to high reflectivity, achieved with reflective paints. The result was a power consumption ranging from about 53,300 kWh/year, for low reflectivity, up to about 39000 kWh/year, for high reflective paints, with a saving of about 2.5% on the current annual energy consumption.

## ACKNOWLEDGMENTS

Discussions with Biagio Morrone, of the DIII of the Second University of Naples are gratefully acknowledged.

We kindly acknowledge Italian mobile telecommunications provider Wind for the use of their experimental sites. This study was partially funded by Regione Campania through European funds (POR Campania FSE 2007-2013) for the "Dottorato in azienda" project.